\documentclass[fleqn,usenatbib]{mnras}
\usepackage{newtxtext,newtxmath}
\usepackage[T1]{fontenc}

\DeclareRobustCommand{\VAN}[3]{#2}
\let\VANthebibliography\thebibliography
\def\thebibliography{\DeclareRobustCommand{\VAN}[3]{##3}\VANthebibliography}

\usepackage{graphicx}	
\usepackage{amsmath}	
\usepackage{gensymb}
\usepackage{color,soul}
\usepackage{threeparttable}

\newcommand{\pcmsq}{\,cm$^{-2}$}	
\newcommand{\pcc}{\,cm$^{-3}$}
\newcommand{\mpcc}{$m_{\rm p}$ \pcc}
\newcommand{\kmps}{\,km s$^{-1}$}
\newcommand{\ergps}{erg s$^{-1}$}
\newcommand{\phz}{Hz$^{-1}$}

\title[Radio filament in ESO 137$-$006]{Origin of the $\sim$150 kpc radio filament in galaxy ESO 137-006}

\author[Alam et al.]{
Toushif Alam$^{1}$\thanks{E-mail: toushif22@iitk.ac.in}, Kunal P. Mooley$^{1,2}$, Kartick C. Sarkar$^{1,3}$
\\
$^{1}$Indian Institute of Technology Kanpur, Kanpur 208016, UP, India\\
$^{2}$California Institute of Technology, Pasadena, CA 91125, USA \\
$^{3}$Raman Research Institute, Bengaluru, Karnataka, India, 560080\\
}

\date{Accepted XXX. Received YYY; in original form ZZZ} 
\pubyear{\the\year{}}

\begin{document}
\label{firstpage}
\pagerange{\pageref{firstpage}--\pageref{lastpage}}
\maketitle

\begin{abstract}
Sensitive wide-field radio surveys have started uncovering many filamentary structures associated with the jets and lobes of radio galaxies, radio relics in galaxy clusters, and tailed galaxies. Although limited theoretical investigations on the origin of the filamentary structures have associated these filaments with astrophysical shocks and interactions with intracluster magneto-ionic media, more quantitative studies are needed to ascertain their precise nature and origin. Recent MeerKAT observations found peculiar filamentary structures (threaded radio structures) joining the lobes of a nearby FRII-like galaxy, ESO 137-006. 
Here we investigate the origin of these "synchrotron threads" to understand if they may be confined magnetically and could arise due to shocks associated with jet activity. 
Through simulation- and theory-based analysis, we find that the dynamical time ($\sim 70$ Myr) associated with the shock front closely matches the estimated synchrotron age ($\sim 130$ Myr) of the threads, thus making the shock origin hypothesis a favorable scenario for this particular filament.
\end{abstract}




\section{Introduction}\label{sec:intro}
Filamentary structures in radio galaxies have been known for a few decades \citep{hines1989,owen2014}. Over the past few years, owing to the high resolution and sensitive images of the radio sky with telescopes such as the LOFAR \citep{vanHaarlem2013} and the MeerKAT \citep{Jonas2016}, the number of radio galaxies found to harbor such structures has increased manyfold \citep{condon2021,Knowles2022,Brienza2022, Ramatsoku2020, Shimwell2022, vanWeeran2021, Rajpurohit2022, Rudnick2022}. 

The filaments have a fairly broad range of morphologies and associations with the radio galaxies, as diverse explanations have also been provided for their origins.
\cite{condon2021} classify filamentary structures in IC 4296, a low-luminosity radio source reminiscent of the FR I morphology, as "threads", "ribbons", and "rings". 
\cite{condon2021} conjecture that the "threads", occurring in the proximity of the main radio jets and having projected lengths of about 50 kpc, arise from helical Kelvin–Helmholtz instabilities. The "ribbons" (smooth regions around the jets) and "rings", which extend on scales of $\sim$100\, kpc, are relics of jets and vortices, respectively, in the outer portions of the lobes and correspond to jet activity that diminished more than 100 Myr ago. Filaments have also been associated with radio relics \citep{owen2014,Bagchi2006, Gennaro2018, Rajpurohit2022}.
\cite{owen2014} and \cite{Rajpurohit2022} suggest that the filamentary relic in A2256 could be due to a complex shock front, where the radio emission may be due to a distribution of electron acceleration efficiencies from an underlying distribution of Mach numbers, magnetic reconnection between large-scale current sheets sitting at the boundaries between magnetic domains, or fluctuations in the magnetic field strength and topology. An alternative explanation is that the filaments are regions containing preexisting relativistic plasma and magnetic fields stretched by the motion of the intracluster medium (ICM; see also \citet{Gopal-Krishna2001, Dominguez-Fernandez2021}). Filamentary structures have been commonly found through MeerKAT and LOFAR surveys of galaxy clusters and galaxy halos, e.g. Abell 1314 \citep{vanWeeran2021}, A3562 \citep{Venturi2022}, A2034 \citep{Shimwell2016}, Perseus \citep{Gendron-Marsolais2020}; A2255 \citep{Botteon2020}, NGC 507 \citep{Brienza2022}. Many of these structures have steep radio spectra, characteristic of particle aging, and some are associated with polarized radio emission at the $\sim$20\% level. The filaments have straight, bent, and curved morphologies and extensions of $\sim$1--10 Mpc. These structures have been suggested to be foreground radio relics/ merger remnants / AGN remnants \citep[e.g.][]{Pizzo2011,Brienza2022}. Our own galaxy, the Milky Way, also harbors different kinds of linear filaments in the disk and galactic plane \citep{Wang2024, Soler2021}.

\cite{Rudnick2022} recently reported filamentary structure in the cluster A194, which is due to the interaction between a magnetic filament, a radio jet, and a dense ICM clump in that cluster. The two prominent magnetic filaments are at least $200$ kpc long; they curve around a bend in the northern jet of the radio galaxy 3C40B and intersect the jet in rotation measure space. 
\cite{Rudnick2022} suggest, based on X-ray absorption, that relativistic particles and magnetic fields contribute significantly to the pressure balance and also evacuate the plasma in a $\sim$35 kpc cylinder encompassing the filamentary radio structure.
There have been other reports of ICM magnetic fields interacting with radio galaxies, albeit not showing pronounced radio filamentary structures \citep{Guidetti2011, Muller2021}.

\subsection{\texorpdfstring{The radio galaxy ESO 137$-$006 and its filamentary structures}{The radio galaxy ESO 137-006 and its filamentary structures}}

ESO 137$-$006 (coordinates 16:15:03.8 $-$60:54:26; J2000), also known as WKK 6269, is a luminous radio galaxy, having luminosity $L_{1.4\,\rm GHz} \simeq 2 \times 10^{32}$ \ergps \phz \citep{Sun2009} lying at the center of the merging galaxy cluster, Norma \citep[Abell 3627 near Great Attractor Region;][]{Abell1989}, located at a comoving distance of 72 Mpc ($z=0.0162$) \citep{Woudt2009}. The galaxy shows a typical FR-II morphology with its radio lobes bending, thought to be due to the motion of the galaxy towards the main cluster \citep{Sakelliou2000}. 

\cite{Ramatsoku2020} presented MeerKAT radio continuum observations, between 1.0--1.4 GHz, that revealed filamentary structures near the central galaxy ESO 137$-$006, which they referred to as "collimated synchrotron threads" (CSTs) on account of their thread-like morphology. These filaments extend to $\sim$10--150 kpc between the radio lobes (see Figure \ref{fig:meerkat-image}). 
The most prominent CST in ESO 137$-$006, hereby called CST1 is a thin straight structure that joins the eastern and western lobes; it has a width of $\sim$1 kpc and a length of $\sim$150 kpc, and a relatively smooth brightness profile with a peak intensity of a few mJy beam$^{-1}$. 
Two other CSTs originate from the same point in the eastern lobe, CST2 one fading away at a length of $\sim$25 kpc from the lobe and another one, called CST3 making a closed loop with a radius of about 60 kpc and reconnecting the Eastern lobe at its far end. Figure~\ref{fig:meerkat-image} shows the CSTs labeled as Filament(CST1), CST2, and CST3.
Smaller CSTs are seen, connecting the Eastern and Western lobes, closer to the base of the jet. These are especially pronounced in the deep super-resolution image of ESO 137$-$006 generated by \cite{Dabbech2022} through Artificial Intelligence techniques. These smaller CSTs are $\sim$20 kpc long and run somewhat parallel to the current path of jet flow and lie several kpc away from it.
A summary of the positions, sizes, and orientations of the most prominent CSTs in ESO 137$-$006 is presented in Table~\ref{tab:cst_summary}.

\begin{table}
\centering
\caption{Properties of the main collimated synchrotron threads (CSTs) in ESO~137--006.}
\resizebox{\columnwidth}{!}{%
\begin{tabular}{lccc}
\hline
Thread & Central Coordinates (J2000) & Size (Length $\times$ Width) [kpc] & PA$^\ast$ [$^\circ$] \\
\hline
CST1 & RA 16$^{\rm h}$15$^{\rm m}$00$^{\rm s}$, Dec $-60^\circ 55'30''$ & $150 \times 1$ & 45 \\
CST2$^\ast$$^\ast$ & RA 16$^{\rm h}$15$^{\rm m}$20$^{\rm s}$, Dec $-60^\circ 57'00''$ & $80 \times 1$ & 120 \\
CST3$^\ast$$^\ast$ & RA 16$^{\rm h}$15$^{\rm m}$50$^{\rm s}$, Dec $-60^\circ 56'12''$ & $60 \times 1$ & 210 \\
\hline
\end{tabular}%
}
\label{tab:cst_summary}
\\[1ex]
\parbox{\columnwidth}{\small $^\ast$PA: Position angle, measured from north through east.}
\parbox{\columnwidth}{\small $^\ast$$^\ast$CST2 and CST3 are curved, and we give their approximate lengths here.}
\end{table}

\noindent \cite{Ramatsoku2020} speculate that these CSTs could arise from the interaction between the magnetic fields of the radio lobes and the magnetoionic ICM or due to filamentary structures originating from one end of a bent radio lobe that somehow connects back into another lobe.

The radio spectra of the CSTs are steep with $\alpha\simeq2$ ($S_\nu \propto\nu^{-\alpha}$). \citep{Ramatsoku2020} suggest, based on their similar spectral-index distributions, that these CSTs all formed at the same time. CSTs may be unique to the conditions and environment of ESO 137$-$006, but if found to be more ubiquitous through galaxy surveys, then this could open up a new science case for the next generation of radio interferometers.
\cite{Muller2021} presents a magnetic field strength and orientation study for the jellyfish galaxy JO206, which has a large tail with spectral index 2, the same as our CST. In that tail, the magnetic field is very high ($>$ 4.1 $\mu$G) and ordered with high fractional polarization $> 50$\%. The magnetic field vectors are aligned with the ionized gas tail direction. They explained this phenomenon as the draping of magnetic fields on the tail. 

\subsection{Spectral Index Variation in ESO 137-006}\label{sec:specindex}
\cite{Ramatsoku2020} present a spectral index map (their Figure 3, top panel) of ESO~137--006 derived from MeerKAT observations at 1030\, MHz and 1398\, MHz. 
While the CSTs exhibit a relatively uniform but steep spectral index ($\alpha\simeq2.0\pm0.3$), indicating that the electrons have undergone significant synchrotron ageing and are no longer receiving fresh injection of energetic electrons, the spectral indices of the lobes of the radio galaxy distinctly show three regions: inner, middle, and outer regions. In the outer regions of the lobes, the radio spectrum is much steeper than for the CSTs, $\alpha\simeq2.5-4$.
These extremely steep-spectrum zones are likely relics or fossil plasmas, no longer energized by the central engine. 
The morphology of the outer lobes suggests that they may originate from aged backflow material or detached ejecta dispersing into the ICM. 
The middle lobe can be defined as having $\alpha\simeq1.5-2.5$, while the inner lobe regions show flatter spectra ($\alpha\simeq 0.5-1.2$; consistent with more recently accelerated electrons). 
The similarity in spectral index between the CSTs and middle lobe regions suggests a comparable synchrotron age.
These observations support a scenario in which the CSTs may have formed during an earlier episode of AGN activity, analogous to the restarted behaviour seen in several radio galaxies (e.g.\ \citealt{Saikia2009, Konar2012, Mooley2016, Mooley2021}) or possibly associated with forward shocks in a younger cocoon, while the central regions remain dynamically active.
We investigate this hypothesis in the current work.

In this paper, we perform a quantitative investigation of two scenarios that may lead to the formation of CSTs in ESO 137$-$006: \textit{one driven by shocks arising from jet activity and another due to preexisting ICM magnetic flux tubes}. Our main area of interest is the largest CST, hereby called CST1. Other subsequent and less prominent CSTs will briefly follow our analysis. The paper is organized as follows. In section 2, we describe the possible origin of CST1 based on analytical and numerical models. In section 3, we incorporate archival data analysis for X-ray and radio data and estimate synchrotron timescales for CSTs. Section 4 presents the final conclusions of this article and discusses the implications of our results.

\begin{figure}
    \centering
    \includegraphics[width=0.54\textwidth, clip=True, trim={0cm 3cm 0cm 2cm}]{figures/RImage_Annotate_newh.png}
    \caption{1.05 GHz MeerKAT image of ESO 137-006 \citep{Dabbech2022, Ramatsoku2020}, smoothened with a 3-$\sigma$ Gaussian filter.
    The central jets and lobes are labeled along with the largest and most prominent filament, CST1, having a length of $\sim$150 kpc. CST2 and CST3 are also annotated.
    The vertical color bar shows the normalized radio intensity (brightness in mJy pixel$^{-1}$), indicating the relative intensity of the threads to the central object.
    Radio data is normalized to avoid artifacts and NaN data; hence, the intensity range may differ from the image shown in \protect\cite{Ramatsoku2020}.}
    \label{fig:meerkat-image}
\end{figure}

 \section{Analyses using Archival Data}
\label{sec:data}

\begin{figure}
    \centering
   \includegraphics[width=0.99\columnwidth, clip=True, trim={0cm 0cm 0cm 0cm}]{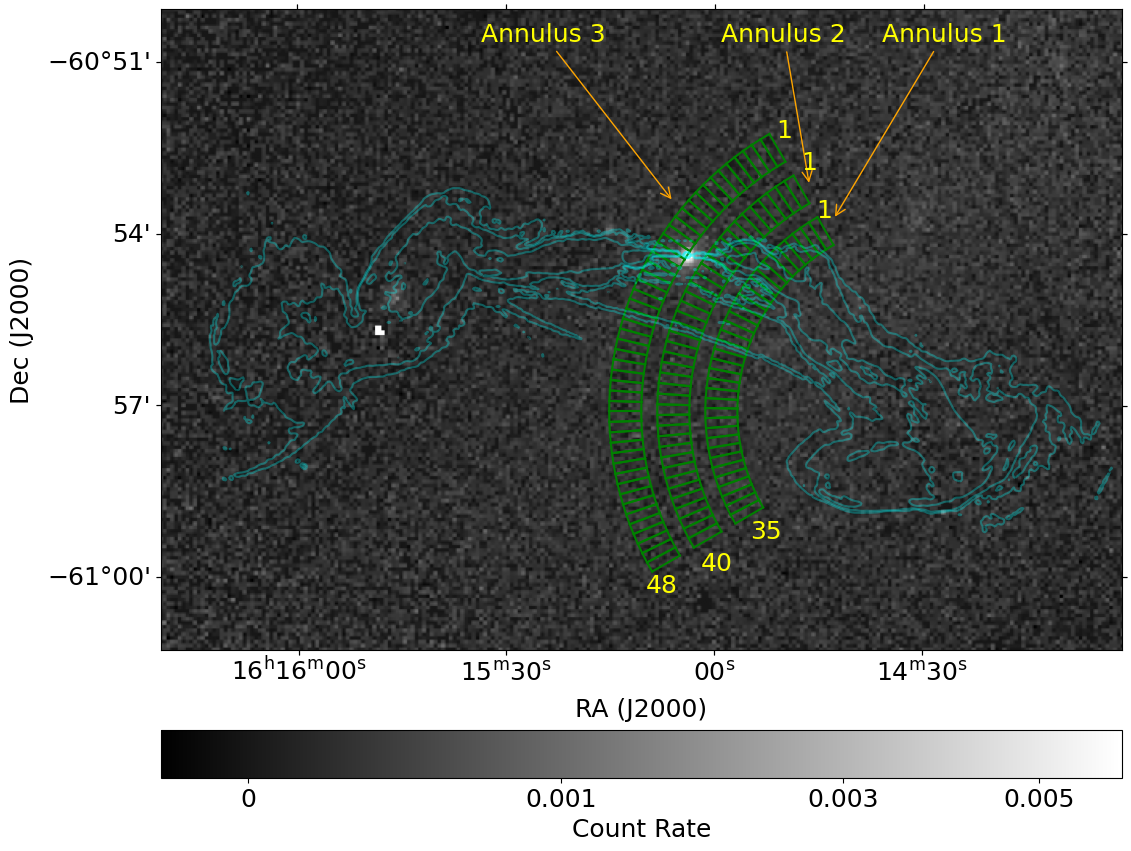}
   \caption{Division of the annular region into sectors over the X-ray image (XMM-Newton XSA repository with Obs ID 0204250101), with radio contours from the 1.05 GHz image having local rms noise of 1.4 $\mu$Jy beam$^{-1}$ areoverlaid (contours are drawn at 0.1, 0.3, 1.3,  4.7, 17.7 mJy beam$^{-1}$ levels). 
   Radio counts are extracted from the maximum value in the central region of the annulus. The horizontal color bar represents the variation in X-ray intensity. Zeroth cells start from the top of the annulus and increase downwards.}
   \label{fig:X-ray-radio}
\end{figure}

To understand the origin of the CSTs, we also investigate the archival radio and X-ray emission from the region.
We use the 1.05 GHz super-resolved MeerKAT image for ESO 137-006 (Project ID SCI-20190418-SM-01) \citep{Ramatsoku2020,Dabbech2022}. The radio intensity map has been shown in Figure \ref{fig:meerkat-image} and as contour lines in Figure \ref{fig:X-ray-radio}.

\begin{figure}
    \centering
    \includegraphics[width=0.99\columnwidth]{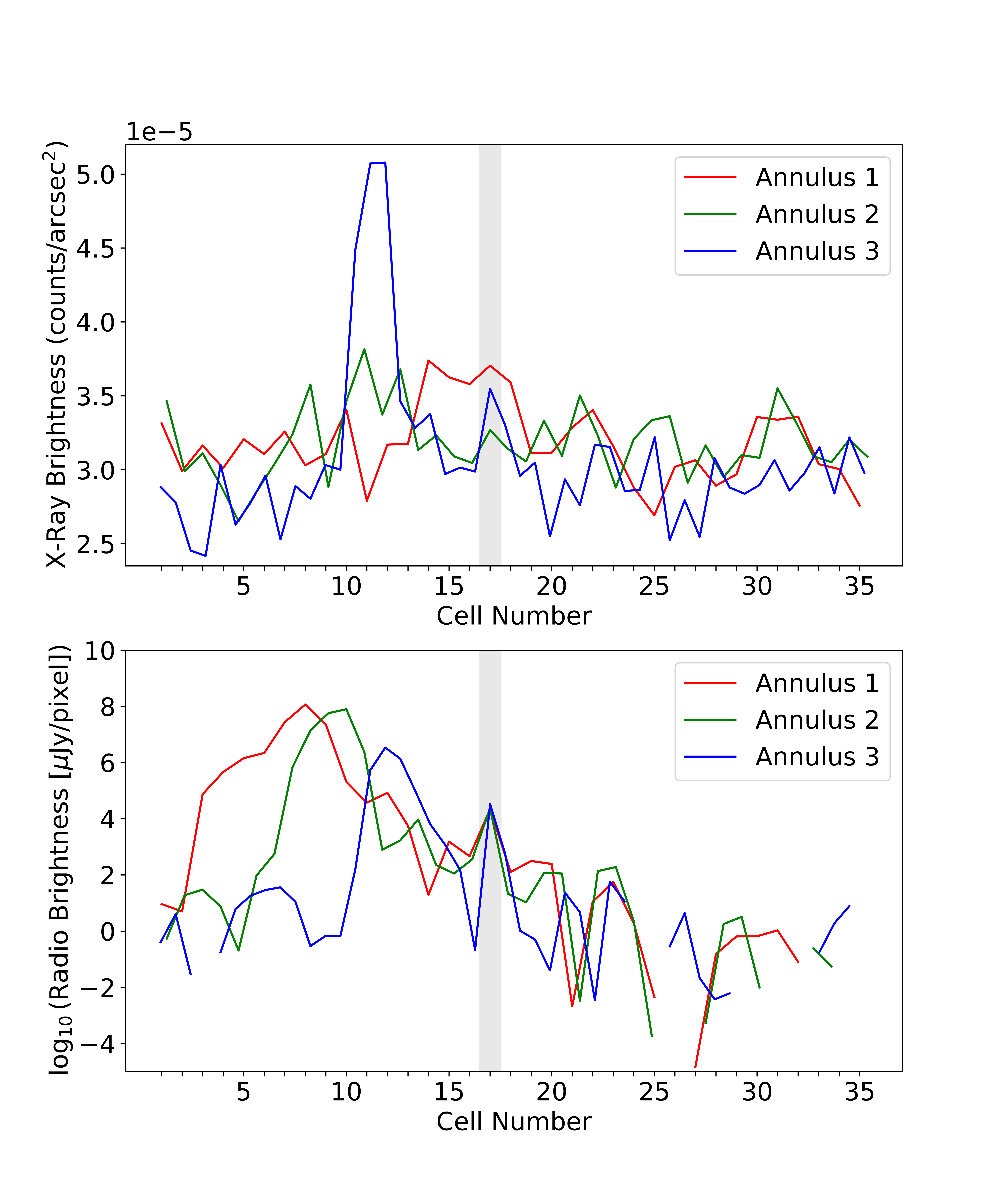}
    \caption{Spatial profiles of photon counts: (a) X-ray surface brightness (2-10 keV) and (b) radio surface brightness in different regions. The grey shaded region marks the cell containing CST1 (the cell numbers are adjusted so that the cell containing CST1 is cell 17 for each of the annuli shown in Figure~\ref{fig:X-ray-radio}. The figure shows that there is no significant change in the X-ray emission in 2-10 keV, with respect to the background, at the location of CST1. At other energy bins (0.2-1 keV and 1-2 keV), X-ray emission is not significant enough to discuss absorption or emission at the location of CST1.} 
    \label{fig:spatial-profiles}
\end{figure}

The ESO 137-006 X-ray data (0.2 to 10 keV) were obtained from the XMM-Newton Science Analysis (XSA) repository (Obs ID 0204250101). The observation file was examined using the standard processing method in Science Analysis Software (SAS) by combining data from the PN, MOS1, and MOS2 detectors. A 3$\sigma$ Gaussian smoothing was applied, and the image was overlaid on radio contours using DS9 to visualize the radio-X-ray morphological correlation (Figure~\ref{fig:X-ray-radio}). 
First, we estimate the surface brightness (SB) of CST1 in the radio image \cite{Dabbech2022} as well as in the X-ray image using features in DS9 tool. 
We find the SB to be $\sim3\times10^{-5}$ counts arcsec$^{-2}$ in the X-ray and $\sim$5\,mJy\,beam$^{-1}$ in the radio. Following \cite{Rudnick2022}, we then investigate the radio and X-ray data to check for X-ray cavities aligned with the threads. We considered only the thread CST1 for this analysis. To understand the spatial correlation of the radio and X-ray emission, we divided the region into three annular regions that intersect CST1 and divided them further into cells.
In Figure~\ref{fig:spatial-profiles}, we plot the X-ray and radio brightness profiles (averaged inside a cell) as functions of cell number. Through this analysis, we do not find any statistically significant evidence for an X-ray cavity or excess emission aligned with CST1. Therefore, inference from the X-ray data remains inconclusive.

\subsection{Synchrotron Age of the CSTs using radio spectral indices}
\label{subsec:synch-age}
The energy loss mechanisms for relativistic electrons, such as synchrotron radiation and inverse Compton scattering, can be described using the Kardashev-Pacholczyk (KP) model \citep{Pacholczyk1970} and the Jaffe-Perola (JP) model \citep{Jaffe1973, Shulevski2015}. The radiative aging of particles in astrophysical environments is explained by these models. The pitch angle, or the angle between a particle's instantaneous velocity and the magnetic field, is assumed to stay constant across time in the KP model. This results in a particular type of anisotropic energy loss. On the other hand, the JP model produces energy losses that are essentially isotropic by taking into consideration the pitch angle's isotropization at time frames that are substantially shorter than the radiation timescale. Either of these models offers a different perspective on electron aging and is appropriate for different physical scenarios. 

For this investigation, the synchrotron age of relativistic electrons at different points along the observed threads was ascertained using a modified version of the JP model, as detailed in \citet{Nandi2021}. 
A crucial piece of information regarding the energy distribution and cooling of the electron population is the spectral index, which is used to compute the synchrotron age. The synchrotron age was specifically calculated using the following equation, which was modified in \citet[][equation 1]{Nandi2021} originally from \citet[][equation 2]{Murgia1999}, for the region where the spectral index $\alpha$ equals 2:
\begin{equation}
    \mathcal{T}_{\rm syn} = 1595\: \mbox{ Myr } \frac{B_{\rm \mu G}^{1/2}}{{B_{\rm \mu G}^2} + {B_{\rm ic, \mu G}^2}} [\nu_{\rm br, GHz}(1+z)]^{-1/2}
\end{equation}
where $B$ is the magnetic‐field strength (in $\mu$G). It is derived from the magnetic energy density, $u_B$, which is further given as a fraction, $\epsilon_B$, of the thermal energy density, i.e.,
$u_{B} = \epsilon_{B}\,u_{\rm gas}$. $B_{\rm ic}$ is the cosmic‐microwave‐background‐equivalent field. The break frequency, $\nu_{\rm br}$, is expressed in GHz. This formulation provides an extensive estimation of the radiative lifetime of electrons by taking into account both inverse Compton and synchrotron losses. SYNAGE software, which is particularly made to calculate synchrotron ages based on spectral index distributions and break frequencies, was used to carry out the calculation. By using this methodology, it is possible to map the age distribution along various thread segments, offering important information on the dynamical and radiative processes that shape the structures that have been observed. Here, $z = 0.016$ stands for the redshift of the source. The filling factor and the ratio of relativistic electron energy density to magnetic energy density dictate the equipartition magnetic field, or $B$.
We consider an intracluster medium (ICM) magnetic field value of 1 $\mu$G \citep{Bonafede2011,Otmianowska-Mazur2003} (refer to case-III in the paper) for age calculation. The magnetic field strength corresponding to the inverse-Compton scattering of cosmic microwave background (CMB) radiation is denoted by the parameter $B_{ic}$. We have 
\[
B_{\rm ic} = 3.2(1+z)^2 \quad \mbox{$\mu$G}
\]
Using the above information, and $B_{\mu G}\sim 1$, we find the synchrotron age from the modified JP model to be $\sim 130$ Myr $\nu_{\rm br, GHz}^{-1/2}$. While we do not expect B$_{\rm ic}$ to be very different from the value considered here, uncertainty in the B$_{\rm \mu G}$ affects the error budget of our calculated synchrotron age. For example, increasing B$_{\rm \mu G}$ by a factor of $5$ decreases the age to $100$ Myr $\nu_{\rm br, GHz}^{-1/2}$. Now, assuming a temperature of $k_{B}T = 6$ keV \citep{Bohringer1996,Nishino2012}, the average sound speed in the ICM is around $c_s = \sqrt{\gamma_a k_{B} T/(\mu m_{p})} \approx 1300$ \kmps. Here we assume $\gamma_a = 5/3$ as the adiabatic index and $\mu = 0.6$ as the mean molecular weight, where $k_B$ is the Boltzmann constant and $m_{p}$ is the proton mass. This estimate for sound speed yields a dynamical age of $\approx$ 150 Myr at a distance of $200$ kpc from the galaxy center.
In the following section, we investigate the possible origin scenarios, for CST1, that are consistent with the above-mentioned observations and estimations.

From the MeerKAT radio spectral map (between $1-1.4$ GHz) of ESO 137-000, we find that the electron power-law index (discussed below) for electrons in CST1 is $p=2\alpha\simeq4$ (synchrotron cooled) or $p=2\alpha+1\simeq3$ (no cooling). 
This is significantly higher than the case of freshly accelerated electrons from diffusive shock acceleration theory \citep{Bell1978} where we expect $p\simeq2$ ($\alpha\simeq0.5)$. 
This suggests that the synchrotron break must be close to the observing frequency of 1.05 GHz.
Additionally, the break frequency cannot be too far below 1 GHz since it would mean the age of the filament is much higher than $\gtrsim200$ Myr, which would likely damage the structural integrity of the filament due to turbulence (see Section~\ref{turbulence}). 
This value of the break frequency is inspired by several other observations of radio lobes. 
Our fiducial value of $\nu_{\rm br}\simeq1$ GHz is also informed by studies of 3C sources in the literature. 
For example, \cite{Mahatma2020} estimated the break frequencies for 3C 320 and 3C 444 to lie between 1- 3 GHz. 
\cite{Turner2018} found the break frequencies of 37 3C sources to range between 100 MHz and several GHz. 
We thus justify our use of $\nu_{\rm br} \sim 1$ GHz.
\footnote{Observations of ESO~137$-$006 below 1\,GHz or above 1.4\,GHz would greatly benefit the inference of the break frequency, but no such useful observations are available. Especially LOFAR/LoTSS (144\,MHz; \citealt{Shimwell2022}), which would be very useful for detecting diffuse emission and filamentary structures at low radio frequencies, does not cover a declination as low as $-60^\circ$.
The Molonglo survey \citep[408\,MHz;][]{Large1981} and SUMSS \citep[843\,MHz;][]{Mauch2003} data exist, but their angular resolutions are too coarse for resolving any filamentary structures.}

\section{Possible Origin of the filament}
\subsection{Thermal/magnetic confinement}
\label{sec:confinement}
There are two main categories of confinement processes that can be invoked to explain the origin of filaments or threads in the ICM: 1) Thermal Pressure Confinement, where the pressure of the filament is dominated by thermal pressure, and 2) Magnetic Field Confinement, where the filament is dominated by magnetic pressure. The creation of such structures, like filaments or threads around a galaxy, is greatly aided by the energy density of cosmic ray electrons and the magnetic field present in the ICM. In this section, our goal is to determine whether thermally or magnetically confined threads are likely in this situation by estimating the corresponding energy densities from synchrotron brightness. If the synchrotron volume emissivity (power per unit volume per unit frequency) for a power‐law electron spectrum $N(\gamma)\,d\gamma = C\,\gamma^{-p}\,d\gamma$ is given as $W_\nu$ then the specific intensity (or brightness) is given as (based on Equation
6.54 of \citet{Rybicki&Lightman}),
\begin{equation}
    S_{\nu} = \frac{W_\nu}{4\pi} l, \quad\,\mbox{\ergps \pcmsq \phz sr$^{-1}$}\,.
\end{equation}
\noindent Here, $l$ is the depth of the emitting region. Following Equation 6.36 of \cite{Rybicki&Lightman}, we write the synchrotron emissivity as 
\begin{eqnarray}
\label{eq:Pnu-RL}
W_{\nu} &=& 3.7\times10^{-23} a(p) B^{(p+1)/2} C\left(\frac{8.4\times10^{6}}{\nu}\right)^{(p-1)/2} \nonumber \\
 && \quad\quad\quad\quad\quad \mbox{erg s}^{-1} \, \mbox{cm}^{-3}  \, \mbox{Hz}^{-1}  \, \mbox{sr}^{-1} 
\end{eqnarray}
where
\begin{eqnarray}
a(p) &=& \frac{\sqrt{\pi} \Gamma\left(\frac{p}{4} + \frac{19}{12}\right) \Gamma\left(\frac{p}{4} - \frac{1}{12}\right) \Gamma\left(\frac{p}{4} + \frac{5}{4}\right)}{2(p+1) \Gamma\left(\frac{p}{4} + \frac{7}{4}\right)} \,,\nonumber \\
\mbox{and } C &=& \frac{u_{\rm cre}(p-2)}{m_{e} c^2} \gamma_{\rm min}^{p-2} \,.
\end{eqnarray}
Here, $N(\gamma) d\gamma = C \gamma^{-p} d\gamma$ represents the electron spectral distribution, $m_{e}$ is the electron mass and $p$ is the electron spectral index. In our calculation, we take the lower cut-off of the Lorentz factor to be $\gamma_{\rm min} = 1$. For simplicity, we assume that the cosmic ray electron energy density, $u_{\rm cre}$, is proportional to the gas energy density, $u_{\rm gas}$, such that $u_{\rm cre} = \epsilon_{\rm cre} u_{\rm gas}$. Here, the thermal energy density is given by $u_{\rm gas} = \frac{n_e k_{\rm B} T}{\gamma_a-1}$, with $\gamma_a = 5/3$ as the adiabatic index and $n_e$ as the electron density. The ICM temperature in the cluster is taken to be $k_{\rm B} T = 6$ keV \citep{Bohringer1996, Nishino2012}. The electron density is taken to be $n_e = 2\times 10^{-3}$ \pcc \citep{Sun2006}. This allows us to compute the internal energy of the thermal gas, $u_{\rm gas} = \frac{3}{2} P$, and the pressure, $P = n_e k_{\rm B} T$. The magnetic energy density, $u_{\rm B} = \frac{\rm B^2}{8\pi}$, can also be found using a similar relation. We assume that the magnetic energy density is proportional to the thermal energy density as $u_{\rm B} = \epsilon_{\rm B} u_{\rm gas}$. We note that the value of $\epsilon_{\rm B}$ decides whether the region is magnetically confined or not. The synchrotron intensity from the thread over the emission depth of $l = 1 \, \text{kpc}$ can be calculated from equations 2--4, which we explain below. Please see Appendix A for more details.

The magnetic and cosmic ray energy densities in equation \ref{eq:Pnu-RL} appear in the form of $\epsilon_{\rm cre} \epsilon_{\rm B}^{(p+1)/4}$, and therefore, are degenerate from each other. To obtain a constraint on these parameters, we following \cite{Kartick2015} and assume $\epsilon_{\rm cre} \epsilon_{\rm B}^{(p+1)/4} = 1$. With this assumption, we compute the synchrotron intensity to be $S_{\nu} = 1.16 \times 10^{10}$ mJy beam$^{-1}$. Here we have used the observed beam dimensions (major axis $= 10$", minor axis $= 9.1$", i.e. 1 beam $=1.68\times10^{-9}$ sr) from \citet{Ramatsoku2020}. On the other hand, 8 mJy beam$^{-1}$ is the value observed at 1.03 GHz. Based on this disparity, we determine that the scaling factor must be
\begin{equation}
\epsilon_{\rm cre}\epsilon_{\rm B}^{\frac{p+1}{4}} = \frac{8}{1.16 \times 10^{10}} = 6.9 \times 10^{-10}.
\label{eq:epsB-epsCR}
\end{equation}
Based on this constraint, we investigate whether CST1 is thermally or magnetically confined. 

\textbf{Case-I: Magnetic confinement}: For magnetic confinement, it is sufficient to assume $\epsilon_{\rm B} = 1$, i.e., half the energy is in the magnetic field (equivalent to $\approx$ 27 $\mu$G). Such a scenario may be applicable to the case of magnetic flux tubes, where the magnetic energy may even dominate the thermal energy density. With this assumption of $\epsilon_{\rm B}=1$, we find $\epsilon_{\rm cre} = 6.9 \times 10^{-10}$. However, we must assess the practical feasibility of such a value for $\epsilon_{\rm cre}$ in the context of the total cosmic ray energy in the ICM. Following \cite{Persic2015}, we use the relation between the cosmic ray electron energy density ($u_{cre}$) and the total cosmic ray energy density ($u_{\rm cr}$) to be
\begin{equation}
   u_{\rm cr} = \left[1 + \left(\frac{m_{p}}{m_{e}}\right)^{\frac{3-p}{2}} \right] u_{\rm cre}.
\end{equation}
We use $p = 2\alpha + 1 = 5$ (where $\alpha=2$)in the non-cooling limit of the electrons. We will discuss the effects of synchrotron cooling in a later section.
This value of $p = 5$ can be substituted into the equation to yield $u_{\rm cr} = 1.0005$ $u_{\rm cre}$. Consequently, the appropriate cosmic ray energy density for $\epsilon_{\rm cre} = 6.9 \times 10^{-10}$ is $\epsilon_{\rm cr} \approx 6.9 \times 10^{-10}$. The energy density of cosmic rays in the ISM typically ranges from $\epsilon_{\rm cr} \sim 0.1$ to $1.0$ \citep{boulares&cox1990}, while in the ICM, simulations produce $\epsilon_{\rm cr} \sim 10^{-5}-10^{-3}$ \citep[see Fig. 12 of][]{Vazza2022}. Our estimated value for the $\epsilon_{\rm cr}$ is excessively low, comparable to the typical ICM values, and seems unreasonable. We, therefore, rule out a magnetically dominated filament for the CST1.

\textbf{Case-II: Cosmic ray dominated $\epsilon_{\rm cr} =1$}:
In this scenario, we assume that cosmic ray energy density is equal to that of thermal gas energy density, which may be applicable to check if there is sufficient magnetic field energy density to form magnetic flux tubes. We use again $\alpha=2$ and $p = 5$. Substituting $\epsilon_{\rm cre} = 0.999 \, \epsilon_{\rm cr} = 0.999$ in equation \ref{eq:epsB-epsCR}, we obtain $\epsilon_{\rm B} = 7.79 \times 10^{-7}$. This is equivalent to a magnetic field strength of 0.024 $\mu$G. The intracluster medium (ICM) magnetic field strength is generally in the range of a few $\mu$G \citep{Bonafede2011,Taylor2002,Govoni2004}. However, three-dimensional numerical simulations conducted by \citet{Otmianowska-Mazur2003} show that the ICM magnetic field strength can be between $\approx 2-5 \, \mu$G. Our estimated magnetic field value is much smaller than the simulated or observed values. 
Therefore, it is doubtful that magnetically confined threads will form under the current conditions.

\textbf{Case-III: Thermal confinement}: As described before, $\epsilon_{\rm cr}$ lie between $10^{-5}$ and $10^{-3}$ in the ICM \citep{Vazza2022}. The same goes again here, $\alpha=2$ and $p = 5$. Using this range for the cosmic ray electrons in equation \ref{eq:epsB-epsCR} produces $\epsilon_{\rm B} \approx 7.8\times10^{-5} - 1.7\times 10^{-4}$. This is equivalent to a magnetic field of $B = 0.2-1.1$ $\mu$G and agrees with the values in ICM but cannot support magnetic field tubes alone (since $\epsilon_B\ll 1$). Therefore, a thermally confined filament is consistent with the observed brightness for the CST1. However, as we show in the next section, the survival of such thermally confined structures in the turbulent ICM may be a challenge.

\subsection{Turbulence in ICM and filament destruction timescale}
\label{turbulence}
The ICM has a relatively low magnetic field strength, and our investigation indicates that the threads' magnetic field strength is $\sim 1$ $\mu$G, much smaller than the thermal energy density (see previous section). In the ICM, the average or root-mean-square (rms) speed of turbulence is usually $\sim 200-300$ \kmps \citep{Vazza2012}. Therefore, a thermally confined filament will be disturbed over a time scale of
\[
t_{\rm dist} = \frac{w_{\rm fil}}{v_{\rm turb}},
\]
where, $w_{\rm fil}$ is the filament width and $v_{\rm turb}$ is the turbulent velocity. Assuming $v_{\rm turb} \sim 200$ \kmps, the destruction time ($t_{\rm dist}$) for a filament with $w_{\rm fil} \sim 1$ kpc is $\sim 5$ Myr. This timescale suggests that turbulence would shatter any thread or filamentary structure within $5\, w_{\rm fil, kpc}$ Myr unless there is any reinforcing mechanisms like shocks or magnetic pressure. For the case of CST1, since the magnetic field is not strong enough ($\epsilon_B \sim 10^{-4}$, B $\sim$ 1 $\mu G$), the other option remaining for such collimated structures to be present even after $\sim 130$ Myr (spectral age) is a shock. The next section discusses a possible model for the shock scenario.

\subsection{Synchrotron threads as shocks}
\label{subsec:shocks}
Shocks traveling through the ICM can maintain a coherent shock front and produce thin radio structures. The shock front produced by an active galactic nucleus (AGN) travels along the direction of jet propagation and also expands orthogonally with time. As the shock propagates, it compresses the magnetic field and accelerates cosmic ray electrons, thus providing an ideal location for synchrotron emission. The width of such structures seen in synchrotron depends on the cooling time of the CR electrons behind the shock. For a cooling time $\mathcal{T}_{\rm syn}$ and a shock speed of $v_s$, the radio filament may appear to have a width of $w \sim v_s\mathcal{T}_{\rm syn}/4$. The factor of $1/4$ accounts for the velocity of the post-shock material with respect to the shock. The actual width of the filament may, however, depend on the cooling rate at a given distance from the shock, which further depends on the electron spectral index \citep{Hoeft2007, Yamasaki2024}. In the rest of this section, we, however, only focus on obtaining the morphological similarity rather than the accurate thickness of the filament, since it requires further investigation into how the CR electron spectrum changes with distance from the shock.

\subsubsection{Bubble Energy}
\label{subsubsec:bubble-energy}

Considering the distance between CST1 and the galactic nucleus, and the height to be about the extent of the lobes, we crudely estimate the dimensions of a cylindrical bubble-like structure carved out by the shock front to have a width of $R_{\rm bub} = 112$ kpc and a length of $2 Z_{\rm bub} = 920$ kpc. Therefore, the total volume is $V_{\rm bub} \approx 4 \times 10^{7}$ kpc$^3$. We further assume that the bubble is expanding with a Mach number $\mathcal{M} \sim 1 $. The temperature of the ICM surrounding ESO 137-006 is taken to be $k_{\rm B} T = 6$ keV \citep{Nishino2012}, or $T \approx 7\times 10^7$ K. The particle number density of gas in the ICM is assumed to be $n \sim 2 \times 10^{-3}$ cm$^{-3}$ \citep{Sun2006, Bohringer1996, Nishino2012}. The pressure of the cluster is then $P = n k_{\rm B} T \approx 2 \times 10^{-11}$ dyne cm$^{-2}$.
Therefore, the total thermal energy contained inside the shock is 
\begin{equation}
    E_{\rm bub, th} = \frac{V_{\rm bub} P}{\gamma_a - 1} \approx 3 \times 10^{61} \mbox{ erg}.
\end{equation}
The total bubble energy would also contain the kinetic energy of the gas. According to the stellar wind bubble evolution model \citep{Weaver1977}, the kinetic energy of the gas inside the bubble is approximately equal to the thermal energy. Therefore, the total energy inside the cylindrical volume of the shock is about $E_{\rm bub} \approx 2 E_{\rm bub, th} \approx 6\times 10^{61}$ erg.

\subsubsection{Calculation of shock propagation}
\label{subsubsec:shock-propagation}
\begin{figure*}
    \centering
    \includegraphics[width=0.99\textwidth]{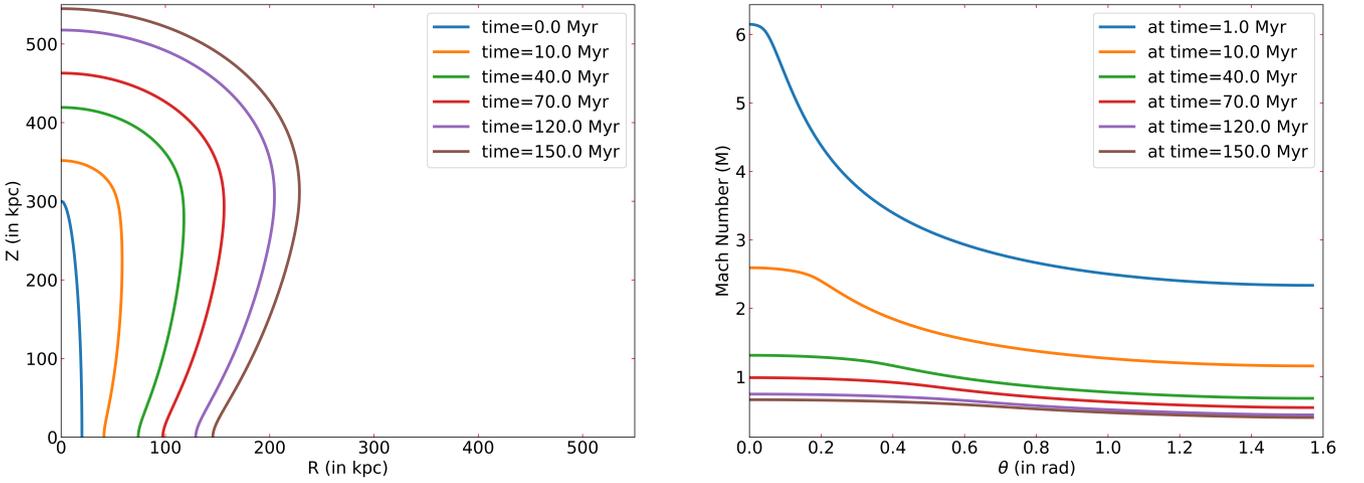}
    \caption{\textit{Left Panel:} Evolution of the shock in AGN and the formation of a bubble for $\beta = 0.75$, $\rho_0 = 0.04$ \mpcc, $r_0 = 10$ \,kpc an initial semi-major axis distance of 300 \,kpc, and an eccentricity of 15. We speculate that the $t = 70$ Myr shock edge (at $Z\lesssim 100$ kpc) roughly mimics the observed radio filament. \textit{Right panel:} Plot of the angle of the shock surface ($\theta$) with the Mach number ($\mathcal{M}$) of the shock at that particular point. Note that $\mathcal{M}<1$ is not physical and only arises mathematically. At this low Mach limit, one has to follow a more accurate calculation of the shock propagation than the one used here.}
    \label{fig:shock-shapes}
\end{figure*}

We assume that the shock is driven by a powerful jet from the central black hole. For simplicity in the theoretical modeling, we further assume that the jet deposited all its energy ($E_{\rm bub}$) in an elliptical region of semi-major axis of $300$ kpc and an eccentricity of $15$. \footnote{We have tested several such initial sizes and shapes and found that this set of initial values produces a final shape that resembles the observed shape of the filament-bubble shape.} This initial shape is shown by $t = 0$ line in figure \ref{fig:shock-shapes}. We solve for the shock propagation in a power-law ICM given by $\rho(r) = \rho_0 (r/r_0)^{-\beta}$.

The evolution of shocks in such a power law medium is difficult to obtain except for a point explosion. For the expansion of an initially non-spherical shock, we follow \cite{Irwin2019} to assume a generalized formalism of Kompaneets Approximation \citep{Komapneets1960}. Following the approximation, we assume that the bubble has a uniform pressure across its entire volume at any given time. Therefore, the pressure, $P$, at any time can be written as 
\begin{equation}
    P(t) = \frac{(\gamma_a-1) E_{\rm bub}}{V_{\rm bub}(t)}\,.
\end{equation}
The shock velocity at any time can then be written as 
\begin{equation}
    v(r, t) = \sqrt{\frac{(\gamma_a+1) P(t)}{2\rho(r)}}
\end{equation}
where $r$ represents the radius at a location on the shock. Now, considering that the shock velocity is always along the direction of the shock surface, we can write down the evolution of a test point on the shock as
\begin{eqnarray}
    R(t+dt) &=& R(t) + v(R, Z,t) \sin\chi \, dt \nonumber \\
    Z(t+dt) &=& Z(t) + v(R, Z,t) \cos\chi \, dt
\end{eqnarray}
where $(R,Z)$ is the coordinate of a test point on the shock surface in cylindrical coordinates and $\chi$ represents the angle between the direction of the surface at that point on the shock and the $Z$ axis. It is given as (see figure \ref{fig:shock-cartoon})
\begin{equation}
    \tan \chi(R, \theta) = \frac{\tan\theta - \frac{d\: \ln R}{d\theta}}{1+\frac{d\: \ln R}{d\theta} \tan\theta}\,
    \label{eq:chi}
\end{equation}
where $\tan\theta = R/Z$. For details, see \cite{Irwin2019}, and see also Appendix~\ref{app:shock-angle} for a demonstration of this calculation. The numerical code has been verified against theoretical expectations for a blast wave and is detailed in Appendix \ref{app:shock-code}.

For the background density, we assume $\rho_0 = 0.04$ \mpcc and $\beta = 0.75$, roughly consistent with the values found in \cite{Sun2006}. The evolution of the shock front for an initial elliptical shock with semi-major axis $=300$ kpc and eccentricity $=15$ has been shown in Figure \ref{fig:shock-shapes}. The figure shows that the initially elongated shock propagates slowly close to $Z=0$ axis compared to the $R=0$ axis. This is because of lower density at $R\sim 300$ kpc compared to $R\sim 20$ kpc. After a certain time, we see that the shock becomes inflated toward the high $Z$ values. It is interesting to note the similarity between the shape of the shock at $t = 70$ Myr and the current size of the radio bubble in ESO 137-006 ($R_{\rm bub} \sim 120$ kpc and $Z_{\rm bub} \sim 460$ kpc). In fact, the shock at $R\sim 100$ kpc is almost parallel to the Z axis, where the AGN jet is supposed to be. This shows a strong possibility that filament CST1 can simply be the forward shock front of the cocoon generated by the AGN in that galaxy.

\subsubsection{Synchrotron Energy Calculation}
\label{subsubsec:shock-synchrotron}

As we see from section \ref{subsec:synch-age} and in the previous subsection, the electron cooling time is comparable to the dynamical time of the filament. Therefore, estimating synchrotron radiation from a shock requires one to also include the effects of CR electron cooling. Such cooling is not included in equation \ref{eq:Pnu-RL}. For this purpose, we make use of Eq. 32 from \citep{Hoeft2007} and calculate the synchrotron radiation behind the shock. We take the typical number density of the ICM to be $n_e \approx 2 \times 10^{-3}$ \pcc, consistent with the values used earlier in this section. We then obtain the result for the emission at $\nu = 1030$ MHz and at $1398$ MHz, as observed through MeerKAT as
\begin{eqnarray}
    \frac{dP\nu}{d\nu}
    &=& 6.4 \times 10^{34} \, \text{erg s}^{-1} \text{Hz}^{-1}\: \frac{A}{\text{Mpc}^2} n_{e,-4} \, \frac{\epsilon_{\rm e}}{0.05}\, \nu_{1.4 {\rm GHz}}^{-p/2} \nonumber \\
   &\times& \left( \frac{T_d}{7 \, \text{keV}} \right)^{3/2} \frac{\left( \frac{B}{\mu G} \right)^{1 + p/2}}{\left( \frac{B_{\text{CMB}}}{\mu G} \right)^2 + \left( \frac{B}{\mu G} \right)^2} \, \Psi(\mathcal{M}) \\
   &=& 1.28 \times 10^{36} \, \text{erg s}^{-1} \text{Hz}^{-1}\: \frac{A}{\text{Mpc}^2} \left( 80 \pi \right)^{\frac{p-2}{4}} n_{e,-4} \\
   &\times& \nu_{1.4 {\rm GHz}}^{-p/2} \, \left( \frac{T_{\rm d}}{7 \, \text{keV}} \right)^{3/2}  \nonumber \,\left(u_{\rm gas,-11} \right)^{\frac{p-2}{4}} \, \epsilon_{\rm e}\, \epsilon_{\rm B}^{\frac{p-2}{4}} \, \Psi(\mathcal{M})
   \label{eq:Pnu-Hoeft07}
\end{eqnarray}
where
$\Psi(\mathcal{M}) = 10^{-4}$, $n_{e,-4} = n_e/(10^{-4} \text{\pcc})$, and $\nu_{1.4GHz} = \nu/1.4$ GHz. $u_{\rm gas,-11}=u_{\rm gas}/(10^{-11})$ ergs cm$^{-3})$. We have further neglected $B_{\text{CMB}}$ (equivalent magnetic field of the CMB energy density and is $\sim 3.35$ $\mu$G, as $B_{\text{CMB,z}}=B_{\text{CMB,0}}(1+z)^2$ and $B_{\text{CMB,0}}=3.24$ $\mu G,z=0.0162$) as it is significantly smaller than $B$, and assumed $\Psi(\mathcal{M})=10^{-4}$ for a weak shock. The area $A$ is considered as the emission region from the thread for a 1 kpc region, i.e. $A=1\times 1$ kpc$^2$, and $T_d = 6$ keV (ICM temperature). For $\nu = 1030$ MHz, at $p=4$, to achieve the observable intensity of 8 mJy beam$^{-1}$, the product $\epsilon_{\rm e}\epsilon_{\rm B}^{(p-2)/4} = 0.00057$ which implies $\epsilon_{\rm e}=0.002$ (for $\epsilon_{\rm B}=0.1$, B = 8.5 $\mu$G). Similarly, for $p=5$, the product $\epsilon_{\rm e}\epsilon_{\rm B}^{(p-2)/4} = 0.00067$, implying $\epsilon_{\rm e}=0.004$ (for $\epsilon_{\rm B}=0.1$, B = 8.5 $\mu$G). All these are reasonable values for the ICM magnetic field and cosmic ray density. The above numbers suggest that there is a strong possibility that these filaments are shocks driven by the jet/cocoon structure.

\begin{figure*}
    \centering
    \includegraphics[width=0.99\textwidth]{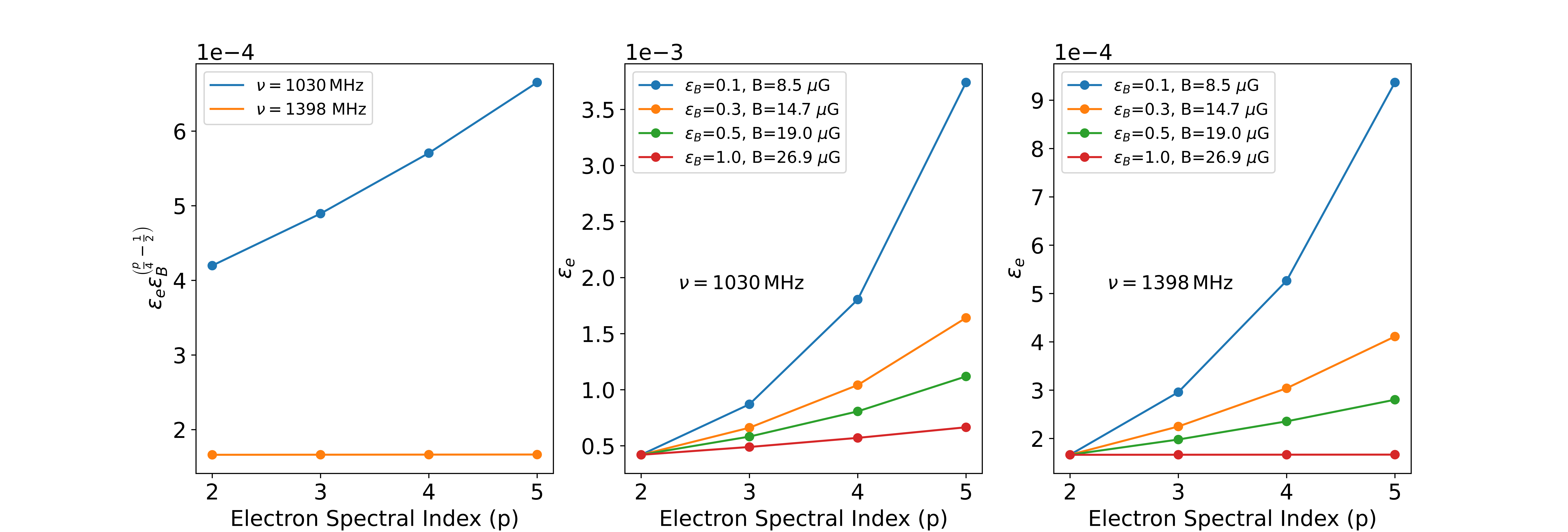}
    \caption{
        Magnetic Field Energy Density and Cosmic Ray Electron Energy Density Fractions and Parameters. 
        \textbf{Left:} Fractional product of cosmic ray electron and magnetic field energy density. 
        \textbf{Middle:} Synchrotron power of 8 mJy beam$^{-1}$ at $\nu=1030$ MHz. 
        \textbf{Right:} Synchrotron power of 4.3 mJy beam$^{-1}$ at $\nu=1398$ MHz.
    }
    \label{fig:energy-fraction}
\end{figure*}

A more detailed analysis, including several values of $p$ and for both the frequency levels, is plotted in Figure \ref{fig:energy-fraction}, where we can see the change and variation of the energy density ratios as a function of $p$.

\section{Discussion}
\label{sec:discussion}

\subsection{Ram pressure stripped Galaxies}
As noted earlier, the radio filaments can also be created due to aligned magnetic fields in the tail of a stripped galaxy \citep{Muller2021}. In such a case, the low temperature ($\sim 10^4$ K) tail could also produce X-ray absorption in addition to the radio emission. This may be the case for the filament analyzed in \cite{Rudnick2022}. In our case, the X-ray surface brightness lacks any characteristic information near CST1 to comment on such a scenario. Further verification can be done if one can track the radio filament with a galaxy at its tip, and it may require one to track the kinematics information of all the nearby galaxies in the cluster. Such a study is out of the scope of the current paper.

\subsection{X-ray signatures of the shock}
While we bat for the shock origin of the CST1, it is difficult to verify such a theory. One possibility would be to detect a jump in X-ray surface brightness at CST1. Such a jump would imply an enhancement of the ICM density, possibly due to a shock. However, as noted in figure \ref{fig:shock-shapes}, the Mach number of the shock is $\mathcal{M}\sim 1$. Such low Mach shocks do not produce significant density compression that can be detectable using modern telescopes. This is especially true when such tiny compression is to be distinguished against the X-ray bright ICM.

Another possibility to detect such a shock would be to detect bulk motion in some intense X-ray lines, such as the $6.6$ keV He-$\alpha$ line of Fe. Such detections are possible using high spectral resolution telescopes such as \textit{XRISM} and \textit{Athena}. However, such bulk motion may be confused with the bulk motions due to ICM turbulence \citep{Xrism2025}.

\subsection{Polarization and Multi-frequency Observations}
For a shock-compressed material, the magnetic field would be parallel to the filament. Therefore, radio polarization information at the CST1 would provide us with another checkpoint for the shock origin of the CST1. 
However, polarization observations are currently unavailable, limiting the ability to fully understand the magnetic field orientation. 
Future observations with more multi-frequency data will also significantly improve the spectral aging analysis and help refine these results.

\subsection{Accretion rate of the central black hole}
Our results allow us to estimate the activity of the central supermassive black hole (SMBH) situated at the center of ESO 137-006 (also known as WKK 6269 galaxy). Considering that the total energy of the AGN cocoon is $E_{\rm bub} \sim 6\times 10^{61}$ erg and that its age $\mathcal{T}_{\rm syn} \sim 130$ Myr, the mechanical power of the jet can be estimated to be $\mathcal{L}\sim E_{\rm bub}/\mathcal{T}_{\rm syn} \approx 1.5\times 10^{46}$ \ergps.

Unfortunately, there is no direct estimation of the SMBH mass for this galaxy. An indirect mass estimation can be done based on the stellar mass of the galaxy. Galaxy ESO 137-006 is one of the cD galaxies in the Norma cluster which has a mass of $M_{\rm norma} \approx 10^{15}$ M$_\odot$ \citep{Ramatsoku2020}. Now, considering that the cD galaxies dominate the mass at $r\lesssim 30$ kpc, they contain about $1$ \% of the cluster mass \citep{Vikhlinin2006}. Therefore, $M_{\rm ESO 137-006} \sim 10^{13}$ M$_\odot$. Now following the $M_{\rm BH}/M_\star \sim 10^{-3}$ relation from \cite{Sarria2010}, we estimate that the SMBH mass in the cD galaxy to be $M_{\rm BH} \sim 10^{10}$ M$_\odot$. This mass corresponds to an Eddington luminosity of $L_{\rm edd} \sim 10^{48}$ \ergps.

The mechanical luminosity for the AGN cocoon is, therefore, $\mathcal{L}/L_{\rm edd} \sim 10^{-2}$. Therefore, the central SMBH seems to be accreting at a sub-Eddington rate. We should, of course, note that this number is uncertain by a factor of $\sim 10$ due to the inherent uncertainty of the $M_{\rm BH}-M_{\rm norma}$ relation used here.

\subsection{Open-loop CSTs}\label{sec:lowfreq}
Filamentary threads inside radio lobes have previously been interpreted as relic jets \citep{Murgia2004,Clarke2004}, but in such a scenario, open-loop CSTs remain unexplained. 
In our AGN‐shock scenario, a forward shock can produce a filament along the jet axis (e.g. CST1 in ESO 137-000). Secondary hydrodynamic instabilities at the contact discontinuity (due to Kelvin–Helmholtz instability) can create partially closed shocks that can lie between the forward shock and the contact discontinuity. Such secondary shocks can generate shorter‐length CSTs that do not form closed loops. These instabilities and partial shocks that arise along with the main shock can reproduce the observed 'broken' morphology of CST2 and CST3 in ESO 137-000 and also several other small filamentary structures seen along its radio lobes. 
Hence, a forward shock scenario can explain the closed-loop CSTs.

\section{SUMMARY \& Conclusion}
\label{conclusion}
In this paper, we have investigated the origin of the most prominent radio filament (collimated synchrotron thread, CST), which spans $\sim 150$ kpc and appears to connect the two lobes of the radio galaxy ESO 137-006. We have used MeerKAT radio (1--1.4 GHz) and XMM-Newton X-ray (0.2-10 keV) maps, together with synchrotron age ($\mathcal{T}_{\rm syn}$) calculation, and solving for forward-shock propagation, to constrain different origin scenarios for CST1.
Below, we list the main findings of our investigation and related discussion. 

\begin{itemize}

\item \textbf{Energy and age:} We estimate that the AGN-driven radio bubble in this galaxy contains an energy of $E_{\rm bub} \sim 6 \times 10^{61}$ erg. 
We also estimate that the age of the filament, based on synchrotron cooling, is $\mathcal{T}_{\rm syn} \sim 130$ Myr (see Section~\ref{subsec:synch-age} for uncertainties introduced due to the ICM magnetic field and the synchrotron break frequency).
We thus find the total mechanical luminosity of the jet to be $\mathcal{L}\sim 10^{46}$ \ergps or $\sim 10^{-2}\, L_{\rm edd}$.

\item \textbf{Thermal/magnetic confinement:} We estimate that a thermally confined magnetic filament can give rise to the radio filament. In such a case, the cosmic ray energy density = magnetic energy density $\ll$ thermal energy density in the filament. In other words, the filament is not in `equipartition'. We further estimate that $B \approx 1\ \mu$G is sufficient to produce the observed radio intensity. However, such structures are dynamically unstable due to the turbulence in the ICM. The turbulent destruction time is $\sim 5$ Myr ($\ll$ the synchrotron age). Therefore, such thermally confined thin filaments could not possibly give rise to the CST1.

\item \textbf{Radio filament, CST1, as a shock:} 
\label{dyn-time}
Based on the similarity of the dynamical time ($\mathcal{T}_{\rm dyn} \sim 70$ Myr) of the shock generated by the AGN cocoon and the synchrotron cooling timescale ($\mathcal{T}_{\rm syn} \sim 130$ Myr), we suggest that the main radio filament CST1 connecting the two radio lobes in the galaxy is most probably represent the forward shock of the AGN cocoon. This is further supported by the morphological similarity of the forward shock and CST1. The resulting radio intensity in such a shock is also consistent with the observed value.

Our investigation, therefore, suggests that many of the filaments in radio galaxies could be weak shocks, either generated due to the turbulent motion at the lobe (where the jet terminates) or the forward shock itself. Validating the above-proposed scenario would require one to correlate the dynamical time of the shock to the synchrotron cooling time and detailed X-ray brightness maps in those regions. 

These results strongly support an AGN‐driven forward‐shock as a viable scenario to explain the origin of CST1, with supplementary hydrodynamic instabilities at the contact discontinuity accounting for the open-loop sub‐structures.  
Future deep observations with the SKA‐low and SKA-mid, together with polarization observations, could reveal fainter substructures in ESO 137-006 and further constrain the properties of the CSTs in this radio galaxy.

\end{itemize}
\section{Acknowledgements}\label{sec:Acknowledgements}
We acknowledge Mohan Karthik G for his key suggestions and help in understanding the simulation studies better.
We thank the anonymous referee for giving valuable comments that helped improve the paper.
We also acknowledge Oleg Smirnov, Mpati Ramatsoku, and Landman Bester, from whom we obtained useful information. 

\section{Data availability}\label{sec:Data availability}
No new data were generated or analysed in support of this research. The MeerKAT data are publicly available via SARAO; the XMM‐Newton data via XSA (ObsID 0204250101).

\bibliographystyle{mnras}
\bibliography{example}


\appendix

\section{Energy density calculation}

\unboldmath
Number density of electrons:
\[
n(E) dE = k E^{-x} dE 
\]
So total number density of electrons throughout the energy range:
\[ N = \int_{E_{\text{min}}}^{E_{\text{max}}} n(E) \, dE = \int_{m_{e} c^2}^{E_{\text{max}}} k E^{-x} \, dE \]

\[ N = k \left[ \frac{E^{-x+1}}{-x+1} \right]_{E_{\text{max}}}^{E_{\text{max}}} = \frac{k}{1-x} \left( E_{\text{max}}^{1-x} - (E_{\text{min}})^{1-x} \right) \]

Now, the energy density of these electrons over all energies will be
\begin{align}
u_{\text{cre}} &= \int_{E_{\text{min}}}^{E_{\text{max}}} E n(E) \, dE 
= \int_{E_{\text{min}}}^{E_{\text{max}}} k E^{-x} E \, dE \nonumber = \int_{m_{e} c^2}^{E_{\text{max}}} k E^{-x+1} \, dE \nonumber \\
&= k \left[ \frac{E^{-x+2}}{-x+2} \right]_{E_{\text{min}}}^{E_{\text{max}}} 
= \frac{k}{2-x} \left( E_{\text{max}}^{2-x} - E_{\text{min}}^{2-x} \right)
\end{align}
\noindent(for \( x \neq 2 \), this expression is valid.)

\[
\therefore \quad k = \frac{u_{cre} (2-x)}{E_{\text{max}}^{2-x} - (E_{\text{min}})^{2-x}} \]

Therefore, the expression for \( k \) in terms of the cosmic ray energy density \( u_{cre} \), the electron spectral index \( x \), and considering the lower cut-off of the Lorentz factor \(\gamma \approx 1\) is:
\[ k = \frac{u_{cre} (2-x)}{E_{\text{max}}^{2-x} - (m_{e} c^2)^{2-x}} \]
\\
Alternatively, if we use the gamma form of electron energy equation: 
\[
N(\gamma) d\gamma = C \gamma^{-p} d\gamma
\]
So total number density of electrons throughout will be:
\[ N = \int_{\gamma_{\text{min}}}^{\gamma_{\text{max}}} n(\gamma) \, d\gamma = \int_{m_{e} c^2}^{\gamma_{\text{max}}} C \gamma^{-p} \, d\gamma \]

\[ N = C \left[ \frac{\gamma^{-p+1}}{-p+1} \right]_{\gamma_{\text{max}}}^{\gamma_{\text{max}}} = \frac{C}{1-p} \left( \gamma_{\text{max}}^{1-p} - (\gamma_{\text{min}})^{1-p} \right) \]

Similarly, the total energy density of these electrons will be
\[ u_{\text{cre}} = \int_{\gamma_{\text{min}}}^{\gamma_{\text{max}}} \gamma m_{e}c^2 n(\gamma) \, d\gamma = \int_{\gamma_{\text{min}}}^{\gamma_{\text{max}}} Cm_{e}c^2 \gamma^{-p} \gamma \, d\gamma \]
\[u_{\text{cre}} = \int_{\gamma_{\text{min}}}^{\gamma_{\text{max}}} Cm_{e}c^2 \gamma^{-p+1} \, d\gamma \]
\[ u_{\text{cre}} = Cm_{e}c^2 \left[ \frac{\gamma^{-p+2}}{-p+2} \right]_{\gamma_{\text{min}}}^{\gamma_{\text{max}}} = m_{e}c^2\frac{C}{2-p} \left( \gamma_{\text{max}}^{2-p} - \gamma_{\text{min}}^{2-p} \right) \]
(for \( p \neq 2 \), this is valid)

\[
\therefore \quad C = \frac{u_{cre} (2-p)}{m_{e}c^2(\gamma_{\text{max}}^{2-p} - \gamma_{\text{min}}^{2-p})} \ = \frac{u_{cre} (p-2) \gamma_{\text{min}}^{p-2}}{m_{e}c^2 \left( 1 - \left(\frac{\gamma_{\text{max}}}{\gamma_{\text{min}}}\right)^{2-p} \right)} \]

Therefore, the expression for \( C \) in terms of the cosmic ray energy density \( u_{cre} \), the electron energy distribution index \( p \), by neglecting the term $\left(\frac{\gamma_{\text{max}}}{\gamma_{\text{min}}}\right)^{2-p}$ (i.e., when $\left(\frac{\gamma_{\text{max}}}{\gamma_{\text{min}}}\right)^{2-p} \ll 1$), and considering the lower cut-off of the Lorentz factor $\gamma_{\text{min}} \approx 1$ is:
\[ C = \frac{u_{cre} (p-2)}{m_{e}c^2} \]

\section{Simulation of General Blast-wave}\label{app:shock-angle}
The equation of an ellipsoid centered at the origin is given by:

\begin{equation}
\frac{x^2}{a^2} + \frac{y^2}{b^2} + \frac{z^2}{c^2} = 1
\end{equation}

In spherical coordinates, the equation of the ellipsoid becomes:

\begin{equation}
r^2 \left( \frac{\sin^2\theta \cos^2\phi}{a^2} + \frac{\sin^2\theta \sin^2\phi}{b^2} + \frac{\cos^2\theta}{c^2} \right) = 1
\end{equation}

The 2D projection of the ellipsoid in the plane (typically for cases where the radial component \(r\) is considered along one axis) results in an elliptical shape. The corresponding equation for the elliptical projection in spherical coordinates is:

\begin{equation}
r^2 \left( \frac{\cos^2\theta}{a^2} + \frac{\sin^2\theta}{b^2} \right) = 1
\end{equation}

This can be solved for \(r\) as follows:

\begin{equation}
r = \frac{ab}{\sqrt{b^2 \cos^2\theta + a^2 \sin^2\theta}}
\end{equation}

Now, suppose \((R, \theta)\) represent the coordinates describing the cocoon surface at the initial time \(t = t_0\) when the jet was choked. The initial shape of the forward shock front can be parameterized by a radial function \(R(\theta)\), which describes the radial distance as a function of the angle \(\theta\).

To calculate the evolution of the shock surface, we define \(\chi(\theta)\) as the angle between the surface normal at \((R, \theta)\) and the axis of the shock. The angle \(\chi(\theta)\) is calculated from the tangent of the angle \(\omega\), which represents the slope of the shock surface. The tangent angle \(\omega\) is derived from the radial distance \(R\) as follows:

\begin{equation}
\tan \omega = -\frac{dR}{ds} \quad \text{where} \quad ds = R d\theta
\end{equation}

This equation gives the rate of change of the radial distance \(R\) along the shock surface as a function of the angle \(\theta\), where \(ds\) represents an infinitesimal arc length along the shock front. From this, we can derive:
\[
\therefore \quad \tan \omega = -\frac{dR}{R d\theta} = -\frac{d \ln R}{d\theta}
\]

Thus, the angle \(\omega\) provides information about the local slope of the shock surface at each point.

Now, we can express the angle \(\chi(\theta)\) as the sum of the angle \(\theta\) and the tangent angle \(\omega\):
\begin{equation}
\chi(\theta) = \theta + \omega
\end{equation}

To relate \(\chi(\theta)\) to the tangent function, we use the following identity for the tangent of the sum of two angles:

\[
\therefore \quad \tan(\chi(\theta)) = \tan(\theta + \omega) = \left( \frac{\tan \theta + \tan \omega}{1 - \tan \omega \tan \theta} \right)
\]

Substituting the expression for \(\omega\), we have:

\begin{equation}
\tan(\chi(\theta)) = \left( \frac{\tan \theta - \frac{d \ln R}{d \theta}}{1 + \frac{d \ln R}{d \theta} \tan \theta} \right)
\end{equation}

Finally, we obtain the angle \(\chi(\theta)\) as:

\begin{equation}
\chi(\theta) = \tan^{-1} \left( \frac{\tan \theta - \frac{d \ln R}{d \theta}}{1 + \frac{d \ln R}{d \theta} \tan \theta} \right)
\end{equation}

This angle \(\chi(\theta)\) represents the direction in which the shock front will propagate, deviating from the radial direction \(\theta\) by an additional angle \(\omega\). 

\begin{figure*}
    \centering\includegraphics[width=0.37\textwidth, clip=True, trim={2cm 0.25cm 1cm 0.67cm}]{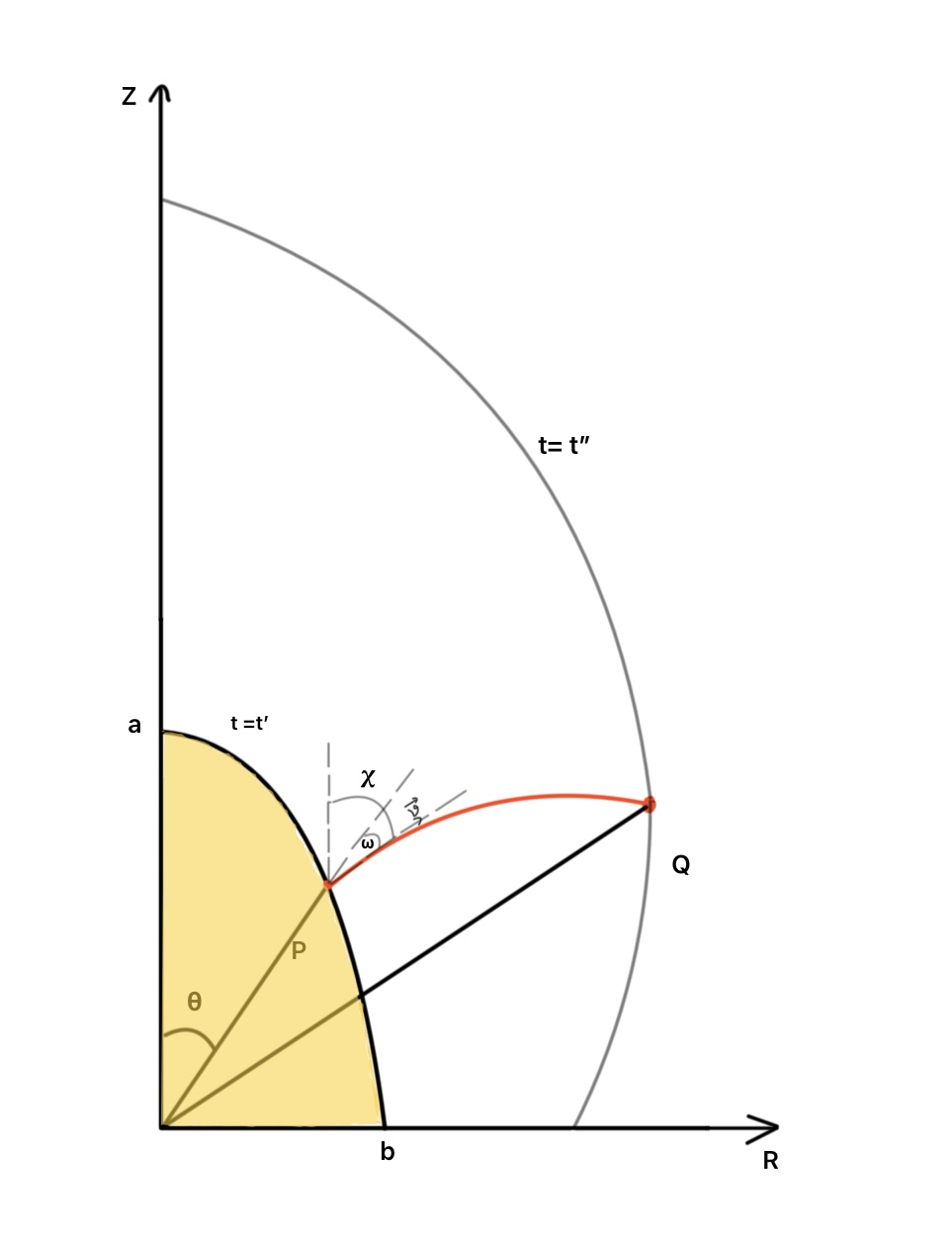}
    \caption{The shockwave propagating into the ambient medium. In a density contrast medium, a particle at point P at the shockfront is making an angle $\chi$ (given in text) and reaches point Q. a and b are the semi-major and semi-minor axes of the initial ellipse, respectively. Angle $\chi$ is greater than $\theta$ by $\omega$ as $\chi=\theta+\omega$ which has been calculated. The time duration is from \(t=t'\) to \(t= t'' \).}
    \label{fig:shock-cartoon}
\end{figure*}

At each point on the surface of the shock, the angle \(\chi\) defines the direction of propagation and is affected by various factors, including the bubble volume, pressure, and ambient density. We incorporate this phenomenology in our code, where we compute the angle \(\chi(\theta)\) at each time step, from \(t = 0\) to \(t = 150 \, \text{Myr}\). By doing so, we can model the evolution of the shock surface and obtain the shape of the filamentary structures, such as the CSTs, which result from the shock interaction with the surrounding medium. A cartoon image Fig.~\ref{fig:shock-cartoon} illustrates the same. The initial elliptical surface is drawn with semi-major axis a and semi-minor axis b. A point P, making an angle $\theta$ with z axis is evolving with velocity $\upsilon$ at the next step making angle $\chi$ and at \(t= t'' \), point P reaches point Q.

\section{Simulation of Spherical Blastwave}
\label{app:shock-code}

Additional information about the spherical shock simulation, the correspondence between analytical and numerical shock propagation, and the geometry for the shock propagation computation—that is, the angles and motion direction—is provided in this section. This phenomenon emphasises how shock-driven processes shaped the thread-like patterns that were seen. For a supernova explosion, a spherical initial surface is assumed due to the isotropic nature of the explosion. In this case, the angle \(\theta\) simplifies, and the shape of the shock front evolves symmetrically, resulting in smooth, spherical bubble shapes. Based on this spherical evolution, we have compared our simulation code with solutions of theoretical equations, and the results match remarkably well, with less than a 1\% error margin. Figure \ref{fig:spheroid-blast-test} illustrates this comparison. The resultant radius of the blast is given by:
\begin{equation}
    r = \sqrt{R^2 + Z^2}.
\end{equation}
We used standard blast wave equations to simulate and analyze this process.
The simulation was made using a custom-built Python code, and the outputs were verified with theoretical predictions, obtaining excellent agreement with an error margin of less than 1\%.
\subsection{Blast Wave Solution}
The following is the expression for the Sedov-Taylor (ST) equation: 
\begin{equation} 
r = \alpha_1 \left(\frac{E_{\rm bub}\,t^2}{{\rho}}\right)^{\frac{1}{5}} \, = \tilde\alpha_1 \left(\frac{E_{\rm bub}}{{\rho_{0}}{r_{0}}^{\beta}}\right)^{\frac{1}{5-\beta}} \, t^{\frac{2}{5-\beta}}, 
\end{equation}

 \noindent where $\rho(r) = \rho_0 (r/r_0)^{-\beta}$ is the density, $E_{\rm bub}$ is the total input energy, r is the blast radius (i.e., the distance the shockfront travels), $\beta$ is a dimensionless blast factor, and $\rho_0$ is the ambient density at $r=r_{0}$,  offers a strong framework for comprehending the evolution of shockfronts, can be used to characterise the conservation of energy in a blast or spherical explosion. 
 A scaling constant that is dependent on the geometry and particular circumstances of the explosion is the parameter $\alpha_{1}$ or $\tilde\alpha_{1}$. Now we already know the bubble energy, ambient density, and $r_0$, which were mentioned in the main section. So using those values and initial condition of radius \(r\) at \(t=t_{0}\) we obtained this $\alpha_{1}$ or $\tilde\alpha_{1}$. Then we take the log of the equation and obtain the equation below.
 \begin{equation}
    \log r = \log \tilde{\alpha}_1 + \frac{1}{5-\beta} \log \left(\frac{E_{\rm bub}}{\rho_{0} r_{0}^{\beta}}\right) + \frac{2}{5-\beta} \log t.
    \label{eq:sol_r}
 \end{equation}

\subsection{Our Code for Simulation}
Every time the shock or blast wave evolves, it contains a specific volume, which is determined in the code by integrating the differential volume, which is considered cylindrical, included in the size of the shock. The shock pressure is computed from the bubble energy injection. We employ the equations for pressure and velocity we have to evolve the surface.
\begin{equation}
    PV = (\gamma - 1)E \quad \quad \upsilon = \sqrt{\frac{(\gamma + 1)P}{2\rho}}
\end{equation}
The initial velocity is $\upsilon_{0}$ for initial pressure $P_{0}$. Then for each point on the surface, we will have
\begin{align}
    R &= R_{0} + \upsilon_{0} \sin\chi \, dt = R_{0} + \upsilon_{0} \sin\theta \, dt, \\
    Z &= Z_{0} + \upsilon_{0} \cos\chi \, dt = Z_{0} + \upsilon_{0} \cos\theta \, dt.
\end{align}
So, the resultant radius of the blast in log form becomes,
\begin{equation}
    \log r = \frac{1}{2} \log \left( R^2 + Z^2 \right)
    \label{eq:code_r}
\end{equation}

\begin{figure*}
 \includegraphics[width=1.0\textwidth, clip=True]{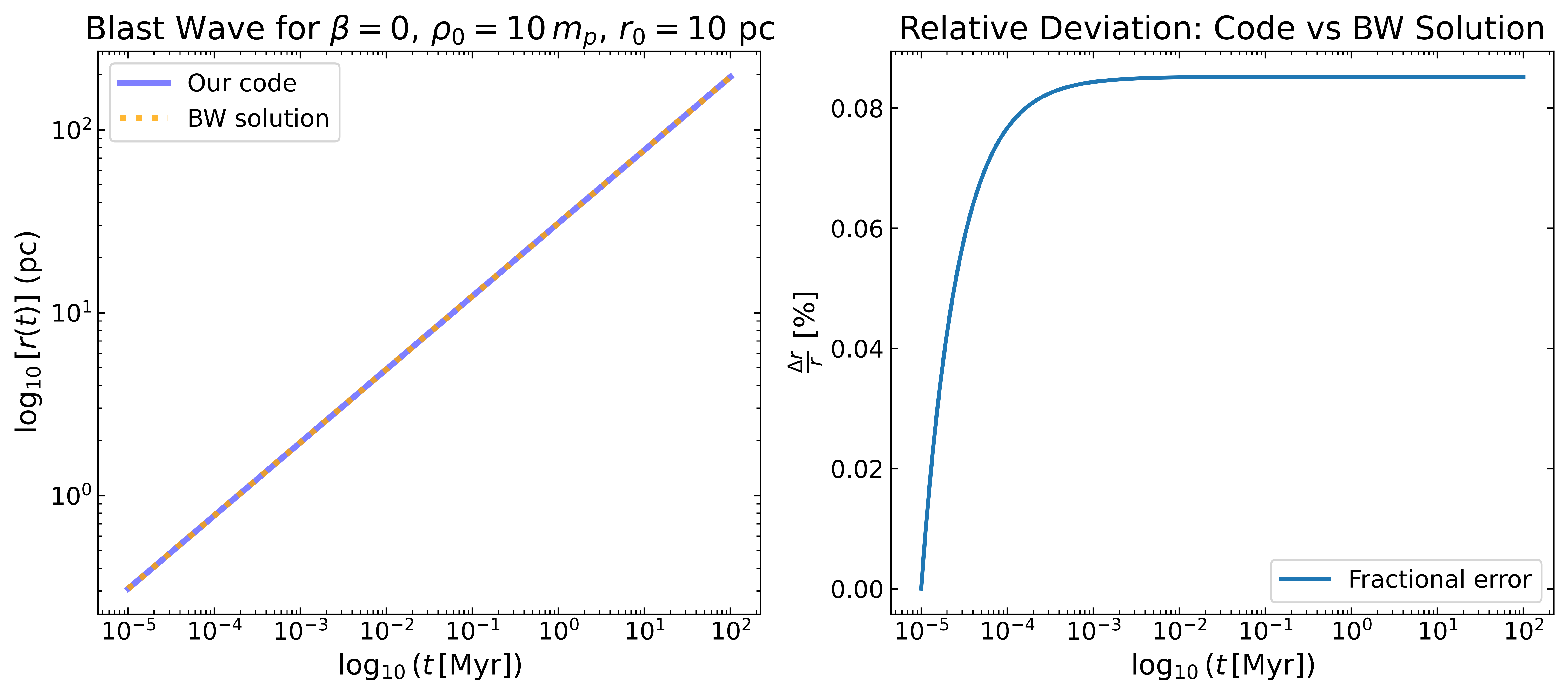}
 \vspace*{-1mm}
 \caption{
 Comparison of the numerical blastwave solution with our analytical result for $\beta = 0$.
 \textbf{Left:} Evolution of the blast radius with time in log--log scale with corresponding legends, showing excellent agreement between the Sedov-Taylor (ST) blastwave solution and our code.
 \textbf{Right:} Percentage error between the ST BW solution and our simulation code, demonstrating minimal deviation and validating the robustness of our implementation.
 }
 \label{fig:spheroid-blast-test}
\end{figure*}

\begin{figure*}
 \includegraphics[width=1.0\textwidth, clip=True]{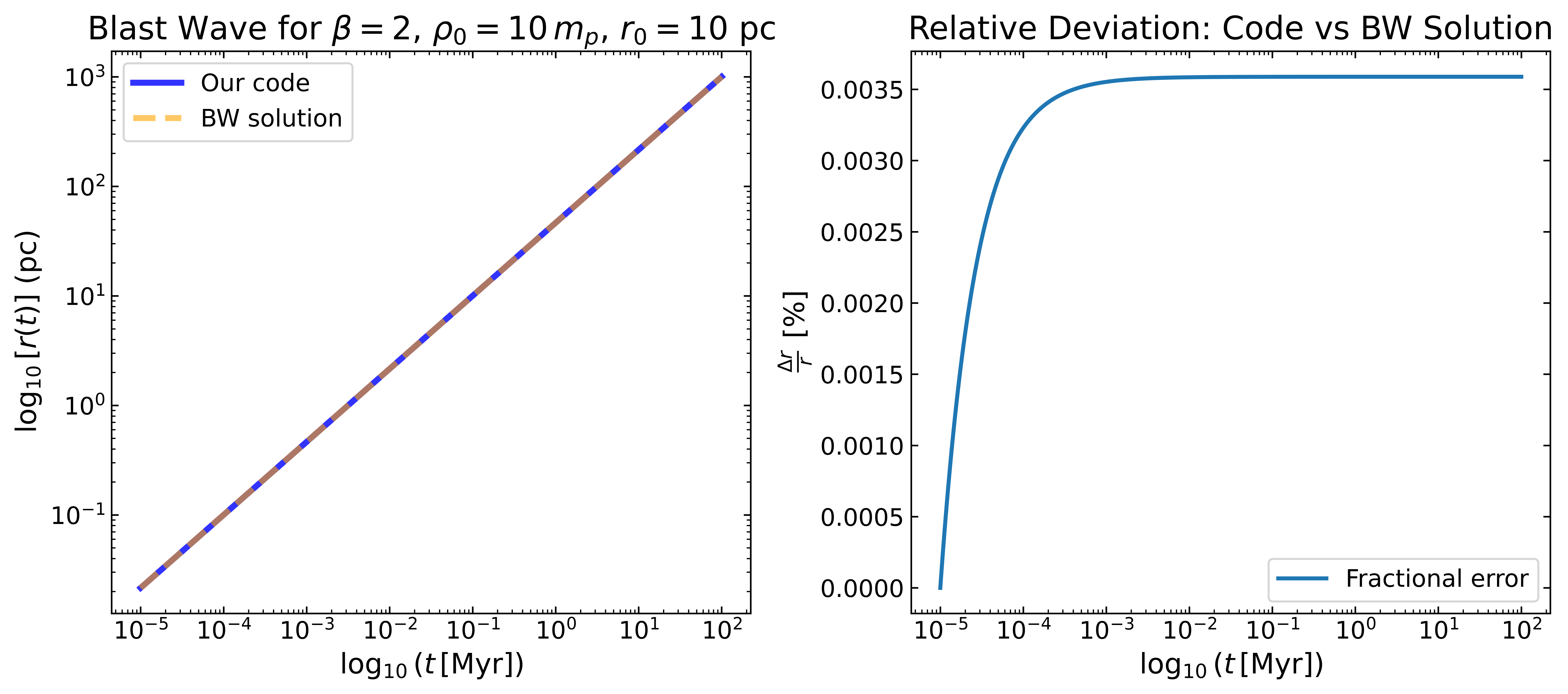}
 \vspace*{-1mm}
 \caption{
 Comparison of numerical blastwave solution with our analytical result for $\beta = 2$.
 \textbf{Left:} Evolution of the blast radius with time in log--log scale for both the blastwave solution and our simulation, with corresponding legends and very close correspondence.
 \textbf{Right:} The relative error in radius over time between the analytical and simulated results, showing consistently low discrepancy.
 }
 \label{fig:blast-wave-test}
\end{figure*}

We compare the blast radius from the numerical blastwave solution (Equation~\ref{eq:sol_r}) and our analytical result (Equation~\ref{eq:code_r}) for $\beta = 0$ (spherical blast) and $\beta = 2$, assuming $\rho_0 = 10\,m_p$, $r_0 = 10$\,pc, and a total injected energy, \(E=E_{bub}=\) $10^{50}$\,erg. The results are shown in Fig.~\ref{fig:spheroid-blast-test} and Fig.~\ref{fig:blast-wave-test}. Our analytical solution is obtained following the formulation given in detail in \citet{Ostriker1988}.

\bsp	
\label{lastpage}
\end{document}